\documentclass[twocolumn,preprintnumbers,amsmath,amssymb,showpacs]{revtex4}
\usepackage{graphicx}% Include figure files
\usepackage{dcolumn}% Align table columns on decimal point
\usepackage{bm}% bold math

\begin{document}

\title{Velocity Profiles in Repulsive Athermal Systems under Shear}

\author{Ning Xu$^1$}
\author{Corey S. O'Hern$^{1,2}$}
\author{Lou Kondic$^3$}
\address{$^1$~Department of Mechanical Engineering, 
Yale University, New Haven, CT 06520-8284.\\
$^2$~Department of Physics, Yale University, New Haven, CT 06520-8120.\\
$^3$~Department of Mathematical Sciences, New Jersey Institute 
of Technology, Newark, NJ  07102.}
\date{\today}

\begin{abstract}
We conduct molecular dynamics simulations of athermal systems
undergoing boundary-driven planar shear flow in two and three spatial
dimensions.  We find that these systems possess nonlinear mean
velocity profiles when the velocity $u$ of the shearing wall exceeds a
critical value $u_c$.  Above $u_c$, we also show that the packing
fraction and mean-square velocity profiles become spatially-dependent
with dilation and enhanced velocity fluctuations near the moving
boundary.  In systems with overdamped dynamics, $u_c$ is only weakly
dependent on packing fraction $\phi$.  However, in systems with
underdamped dynamics, $u_c$ is set by the speed of shear waves in the
material and tends to zero as $\phi$ approaches $\phi_c$.  In the
small damping limit, $\phi_c$ approaches values for random
close-packing obtained in systems at zero temperature.  For
underdamped systems with $\phi<\phi_c$, $u_c$ is zero and thus they
possess nonlinear velocity profiles at any nonzero boundary velocity.
\end{abstract}

\pacs{83.10.Rs,%Computer simulation of molecular and particle dynamics
83.50.Ax,%Steady shear flows
45.70.Mg,%Granular Flow: mixing,segregation,stratification 
64.70.Pf%Glass transitions
}
\maketitle

Driven, dissipative systems are ubiquitous in nature (occurring much
more frequently than equilibrium thermal systems) and display complex
behaviors such as hysteretic and spatially-dependent flows.  Moreover,
there is no complete theoretical description for these systems, which
makes it difficult to predict how they will respond to applied loads
such as shear stress.  Many of these systems such as granular
materials \cite{mueth,granular}, metallic glasses \cite{metallic}, and
complex fluids, for example emulsions \cite{coussot}, foams
\cite{debregeas,dennin}, and worm-like micelles \cite{salmon}, do not
flow homogeneously with a linear velocity profile when they are
sheared.  Shear localization or banding can occur where a small
fraction of the system near one of the boundaries undergoes strong
shear flow while the remainder of the system is nearly static.
Despite much intense work, a complete description of how these systems
respond to shear stress is not available.  We perform molecular
dynamics simulations of repulsive athermal particulate systems in two
(2D) and three (3D) spatial dimensions undergoing boundary-driven
planar shear flow to study mechanisms that give rise to spatially
inhomogeneous velocity profiles.

%%%%%%%%%%%%%%%%%
\begin{figure}
\scalebox{0.5}{\includegraphics{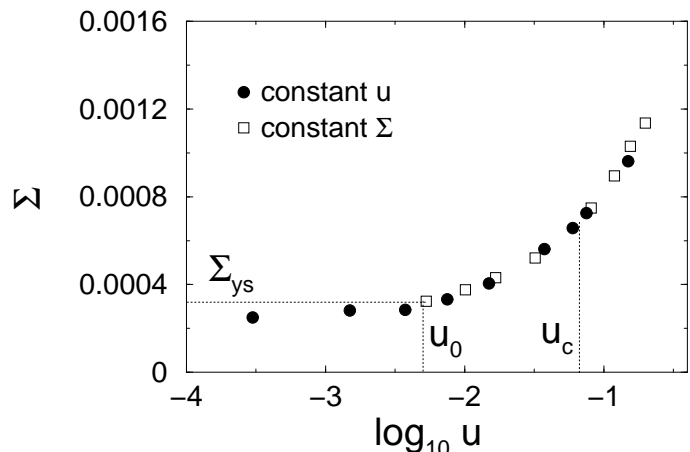}}%
\vspace{-0.25in}
%LK modified; please check
\caption{Shear stress $\Sigma$ vs. velocity $u$ of the wall moving at
constant velocity (circles) and constant stress (squares) for a 2D
underdamped system with linear spring interactions at
$\phi=0.85$. $\Sigma_{ys}$ is the yield stress at constant stress.
Previous studies~\cite{varnik,rottler} have shown that mean
velocity profiles can switch from nonlinear to linear as $u$ increases
above $u_0$, where $u_0$ is the wall velocity at $\Sigma=\Sigma_{ys}$.
We will show below that the mean velocity profiles become nonlinear
again when $u > u_c$, where $u_c \ge u_0$ is always satisfied.}
\label{flowcurve} 
\vspace{-0.3in} 
\end{figure}
%%%%%%%%%%%%%%%%%

We will answer several important questions in this letter.  First,
does the packing fraction of the system strongly influence the shape of
the velocity profiles?  Most previous simulations investigating velocity
profiles in sheared systems have been performed either near random
close-packing as in simulations of granular materials \cite{thompson}
or at high density as in studies of Lennard-Jones liquids
\cite{liem} and glasses\cite{varnik,rottler}.  However, a systematic
study of the role of density has not been performed.  Nonlinear mean
velocity profiles have been found at both high density and near random
close packing, but it is not clear whether the same physical mechanism
is responsible in both regimes.

%LK rewrite
We also consider the influence of the speed $u$ of the shearing
boundary on the mean velocity profiles.  Results from previous
simulations of glassy systems~\cite{varnik,rottler} indicate that a
critical velocity $u_0$ exists below which the mean velocity profiles
become nonlinear \cite{remark2}.  In these systems, the yield stress
$\Sigma_{ys}$ at constant shear stress (shear stress required to move
the boundary at nonzero velocity) is larger than the yield stress
$\Sigma_{yv}$ at constant velocity (shear stress in the limit of zero
velocity).  When the shear stress is between these two values, part of
the system can flow while other parts remain nearly static. In this
letter, we concentrate instead on the larger $u$ regime, and ask
whether the velocity profiles remain linear for all $u > u_0$ .  We
will show that another transition takes place---the velocity profile
switches from linear to nonlinear---when the boundary velocity exceeds
$u_c \ge u_0$.  The onset of nonlinear velocity profiles at large $u$
coincides with the appearance of nonuniform packing fraction and
temperature profiles.  Both $u$ regimes are depicted in
Fig.~\ref{flowcurve} using the flow curve for an underdamped athermal
system in 2D.
 
%%%%%%%%%%%%%%%%
\begin{figure}
\scalebox{0.5}{\includegraphics{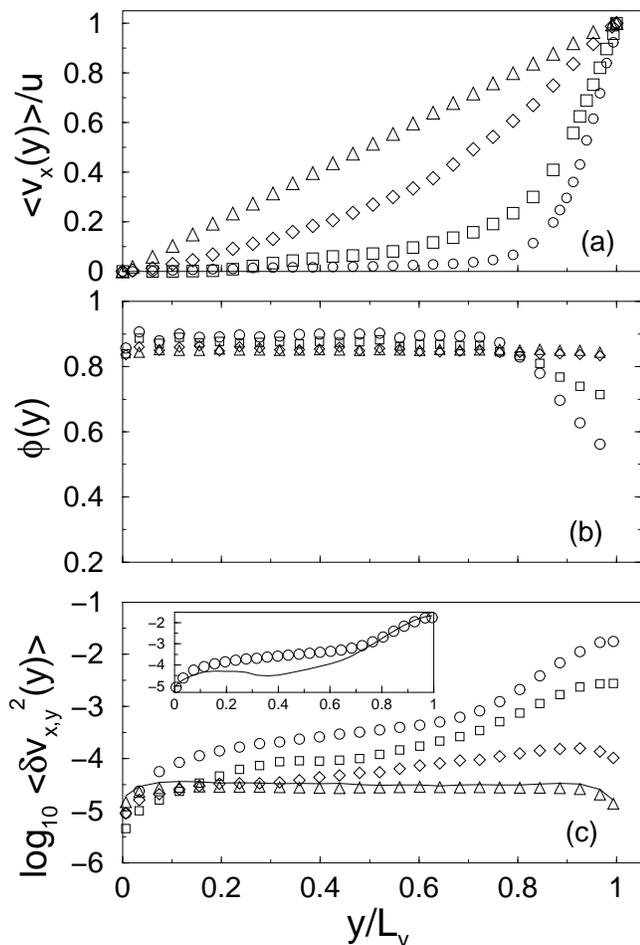}}%
\vspace{-0.25in}
\caption{(a) Average velocity $\langle v_x \rangle$ (normalized by
$u$) in the flow direction, (b) local packing fraction $\phi$,
and (c) velocity fluctuations $\langle \delta v_{x,y}^2
\rangle$ in the $x$- (solid lines) and $y$-directions (symbols) as a
function of height $y/L_y$ from the stationary wall in a 2D
system with harmonic spring interactions and underdamped dynamics
$(b^*=0.01)$ at $\phi=0.85$.  In each panel, $4$ boundary velocities
are shown; triangles, diamonds, squares, and circles correspond to $u
= 0.075$, $0.15$, $0.37$, and $0.75$, respectively.  The inset to (c) 
compares velocity fluctuations in the $x$- and $y$-directions at 
$u=0.75$.}
\label{figure1} 
\vspace{-0.32in} 
\end{figure}
%%%%%%%%%%%%%%%%%
     
In order to demonstrate these results, we performed a series of
molecular dynamics simulations of soft repulsive athermal systems undergoing
boundary-driven shear flow under conditions of fixed volume, number of
particles $N$, and velocity of the top shearing wall $u$.  The systems
were composed of $N/2$ large particles and $N/2$ small particles with
equal mass $m$ and diameter ratio $1.4$ to prevent crystallization and
segregation.  Initial states were prepared by quenching the system
from random initial positions to zero temperature \cite{ohern} using
the conjugate gradient method \cite{numrec} to minimize the system's
total potential energy.  During the quench, periodic boundary
conditions were implemented in all directions.  Following the quench,
particles with $y$-coordinates $y>L_y$ ($y<0$) were chosen to comprise
the top (bottom) boundary.  The walls were therefore rough and
amorphous.  Results did not depend on the thermal quench rate provided
the systems were sheared long enough to remove initial transients.

Shear flow in the $x$-direction with a shear gradient in the
$y$-direction was created by moving all particles in the top wall at
fixed velocity $u$ in the $x$-direction relative to the stationary
bottom wall.  Therefore, particles in the walls do not possess
velocity fluctuations.  During the shear flow, periodic boundary
conditions were imposed in the $x$- and $z$-directions (in 3D).  The
system-size was varied in the range $N=[256,3072]$ to assess
finite-size effects.  Only small sample sizes were required in the $x$
and $z$ directions.  In contrast, more than $\approx 50$ particle
layers were required in the shear-gradient direction to remove
finite-size effects.  Most simulations were carried out using $L_x =
L_z = 18 \sigma$ and $L_y = 72 \sigma$, where $\sigma$ is the small
particle diameter.  The systems were sheared for a strain of $5$ to
remove initial transients and then quantities like velocity, pressure
and shear stress (obtained from the microscopic pressure tensor
\cite{evans}), and local packing fraction were measured as a function
of distance $y$ from the stationary wall.  Averaged quantities were
obtained by sampling between strains of $5$ to $10$.

%LK volume changed to packing below
Bulk and boundary particles interact via the following pairwise,
finite-range, purely repulsive potential: $V(r_{ij}) = \epsilon
\left(1-r_{ij}/\sigma_{ij}\right)^{\alpha}/ \alpha$, where
$\alpha=2$,~$5/2$ correspond to harmonic and Hertzian spring
interactions, $\epsilon$ is the characteristic energy scale of the
interaction, $\sigma_{ij}=(\sigma_i + \sigma_j)/2$ is the average
diameter of particles $i$ and $j$, and $r_{ij}$ is their separation.
The interaction potential is zero when $r_{ij} \ge \sigma_{ij}$.  Our
results were obtained over a range of packing fraction from
$\phi=[0.58,0.80]$ in 3D and $\phi=[0.75,1.0]$ in 2D, which allows us
to probe packing fractions both above and below random close-packing
\cite{ohern}.  The units of length, energy, and time are $\sigma$,
$\epsilon$, and $\sigma \sqrt{m/\epsilon}$, respectively.

For athermal or dissipative dynamics, the position and
velocity of each particle are obtained by solving \cite{luding}
\begin{equation} 
\label{dissipative} 
m \frac{d^2{\vec r}_i}{dt^2} = {\vec F}^r_i - b \sum_j ({\vec v}_{i} 
- {\vec v}_{j}), 
\end{equation} 
where ${\vec F}^r_i=-\sum_j dV(r_{ij})/dr_{ij} {\hat r}_{ij}$, the
sums over $j$ only include particles that overlap $i$, ${\vec v}_{i}$
is the velocity of particle $i$, and $b>0$ is the damping coefficient.
Although for the present discussion we neglect the particles'
rotational degrees of freedom, we have shown that including these does
not qualitatively change any of our results \cite{ning}.  The dynamics
can be changed from underdamped to overdamped by tuning the
dimensionless damping coefficient $b^* = b \sigma/\sqrt{\epsilon m}$
above $b^*_c = \sqrt{2}$.  Frictionless granular materials and model
foams can be studied using $b^* < b^*_c$ \cite{luding} and $b^* \gg
b^*_c$ \cite{durian}, respectively.

%%%%%%%%%%%%%%
\begin{figure}
\scalebox{0.4}{\includegraphics{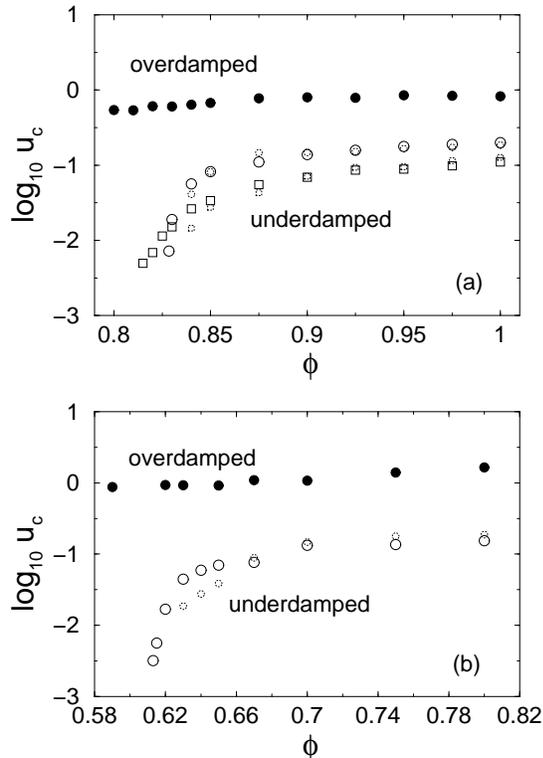}}%
\vspace{-0.23in}
\caption{Critical velocity $u_c$ of the moving wall versus packing
fraction $\phi$ in (a) 2D and (b) 3D systems with harmonic
(circles) and Hertzian (squares) spring interactions. The open and
filled symbols correspond to $b^*=0.01$ and $b^*=5$, respectively. For
underdamped systems, we plot $u_T/2$ for harmonic (small circles)
and Hertzian (small squares) spring interactions, where $u_T$ is the
shear wave speed.}
\label{figure2} 
\vspace{-0.25in} 
\end{figure}
%%%%%%%%%%%%%%%%%

Three physical parameters, the packing fraction $\phi$, the velocity
$u$ of the moving boundary, and the dimensionless damping coefficient
$b^*$, strongly influence the shape of the mean velocity profile.
First, we find that a critical boundary velocity $u_c$ exists that
separates linear from nonlinear flow behavior.  For $u<u_c$ (but not
in the quasistatic flow regime), the mean velocity profiles in the
flow direction are linear; however, when $u>u_c$ they become
nonlinear.  The width of the shearing region decreases as $u$
continues to increase above $u_c$.  This is shown in
Fig.~\ref{figure1}(a) for an underdamped ($b^* \ll b^*_c$) system in 2D
with harmonic spring interactions at $\phi=0.85$.  As $u$ is increased
above $u_c \approx 0.08$, the mean velocity profiles $\langle v_x(y)
\rangle$ become more and more nonlinear.  When the boundary velocity
has increased to $u=0.75$, approximately $80\%$ of the system is
nearly static, while the remaining $20\%$ undergoes shear flow.  

We also monitored the local packing fraction and mean-square velocity
fluctuations (or kinetic temperature) during shear.  These are shown
for the same dense system with underdamped dynamics in
Fig.~\ref{figure1} (b) and (c).  We find that when the mean velocity
profile is linear, the packing fraction and velocity fluctuations are
spatially uniform.  Moreover, the velocity fluctuations in the $x$-
and $y$-directions are identical.  However, when the boundary velocity
exceeds $u_c$, the packing fraction and mean-square velocity profiles become
spatially dependent.  In this regime, the compressional forces induced
by the shearing boundary are large enough to cause dilatancy.  The
system becomes less dense near the shearing wall and more compact in
the nearly static region.  In addition, the shearing wall induces a
kinetic temperature gradient with velocity fluctuations larger near
the shearing boundary. The kinetic temperature also becomes
anisotropic with $\langle \delta v_x^2 \rangle < \langle \delta v_y^2
\rangle$ when $u>u_c$.  Thus, several phenomena occur simultaneously
as the boundary velocity is increased above $u_c$: 1) the velocity
profile becomes nonlinear, 2) the system dilates near the shearing
boundary and compacts in the bulk, and 3) the kinetic temperature
becomes higher near the shearing wall.
          
%LK I thought it would be appropriate to include reference to `remark2'
%at then of this paragraph as well.

We have measured the critical wall velocity $u_c$ as a function of
packing fraction $\phi$ for systems with underdamped dynamics and
harmonic and Hertzian spring interactions in 2D and 3D.  These
measurements are shown in Fig.~\ref{figure2}.  We find that $u_c$ is
nearly constant at large $\phi$ but then decreases sharply as $\phi$
approaches a critical packing fraction $\phi_c$.  For $\phi < \phi_c$,
$u_c=0$ with $\phi_c \approx 0.81-0.82$ in 2D for harmonic and
Hertzian springs and $\phi_c \approx 0.61$ in 3D for harmonic springs.
We expect that Hertzian springs will give a similar result for
$\phi_c$ in 3D.  These values for $\phi_c$ are close to recent
measurements of random close-packing in systems at zero
temperature~\cite{ohern}.  We have also measured $u_0$ in underdamped
systems \cite{ning}.  $u_0$ decreases strongly near random
close packing, but for all systems studied $u_0 \le u_c$.  In
particular, we find that when $u_c=0$, $u_0=0$ also.

A possible interpretation of the critical wall velocity $u_c$ can be
obtained by comparing the time it takes the system to shear a unit
strain to the time it takes a shear wave (with speed $u_T$) to
traverse the system and return to the shearing boundary.  This simple
argument predicts $u_c = u_T/2$.  $u_T$ can be obtained by studying
the transverse current correlation function $C_{T}(\omega,k)$ as a
function of frequency $\omega$ and wavenumber $k=2\pi n \sigma/L_x$
($n=$integer) and the resulting dispersion relation $\omega_{T}(k)$
\cite{hansen}.  In Fig.~\ref{figure2}, we compare $u_{T}=
d\omega_{T}/dk$ (for $n=3$ to $12$) and $u_c$ as a function of $\phi$
for both potentials in 2D and for harmonic springs in 3D
\cite{remark}.  Although deviations occur close to $\phi_c$, we find
that $u_c$ agrees very well with $u_T/2$ over a wide range of $\phi$.

\begin{figure}
\scalebox{0.45}{\includegraphics{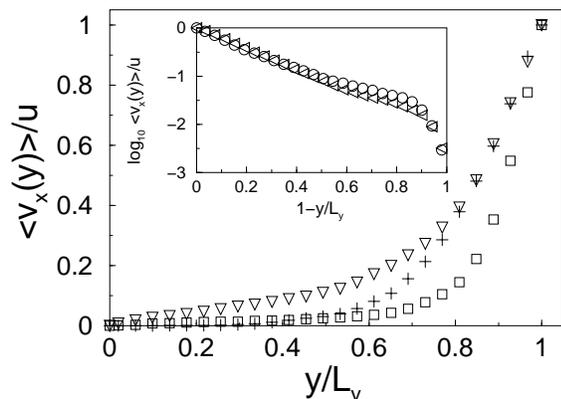}}%
\vspace{-0.26in}
\caption{Average velocity $\langle v_x \rangle/u$ in the shear flow
direction as a function of height $y/L_y$ from the stationary wall in a
2D underdamped system ($b^*=0.01$) at $\phi <
\phi_c$.  Three boundary velocities are shown; squares, downward
triangles, and pluses correspond to $u = 0.38$, $0.038$, and $7.7
\times 10^{-4}$, respectively.  The inset shows that there is a wide
range of $u$ from $0.0077$ (leftward triangles) to $0.077$ (circles)
over which the velocity profiles collapse.}
\label{figure3} 
\vspace{-0.25in} 
\end{figure}

%LK changed last sentence (density -> packing)
What is the shape of mean velocity profiles in dilute underdamped
systems with $\phi < \phi_c$?  Since $u_c$ is zero below $\phi_c$, we
expect that mean velocity profiles in these dilute systems are
nonlinear for all nonzero $u$.  This is indeed what we find for all
systems studied.  Fig.~\ref{figure3} shows the mean velocity profiles
for a 2D underdamped system at $\phi < \phi_c$ over three decades in
$u$.  In contrast to the behavior in dense systems, the velocity
profiles are not monotonic in $u$.  However, there is a range of
boundary velocities (one decade) over which the velocity profiles
collapse onto a common exponential profile.  A robust exponential
profile has also been found over a wide range of shear rates in
experiments of granular materials \cite{granular}.  Similarly to the
systems characterized by large $\phi$, low $\phi$ ones are accompanied
by spatially-dependent packing fraction and mean-square velocity
profiles.

Boundary-driven shear flow in {\it overdamped} systems is, however,
substantially different from that in underdamped systems since
velocities of neighboring particles are strongly coupled.
Fig.~\ref{figure2} shows that in the overdamped limit ($b^* \gg
b^*_c$), the critical boundary velocity is nearly independent of
$\phi$ over the studied range in both 2D and 3D.  We also find that
$u_c$ increases linearly with $b^*$, thus the velocity profiles tend
toward linear profiles as the dissipation increases at fixed $u$.

%%%%%%%%%%%%%%%%%
%%%%%%%%%%%%%%%%%

%LK just to avoid somebody being picky about grammar, perhaps it's
%safer to keep these first sentences in present.  If you don't like
%it, feel free to change it back.

In this letter we present results of molecular dynamics simulations
of repulsive athermal systems undergoing boundary-driven shear flow in
2D and 3D.  We demonstrate that a critical boundary velocity $u_c$
exists (at large $u$) that signals the onset of spatial inhomogeneity.
When $u$ exceeds $u_c$, the mean velocity profiles become nonlinear,
the system becomes dilated near the moving wall and compressed near
the stationary wall, and the system possesses a nonuniform kinetic
temperature profile with higher temperature near the moving wall.  For
underdamped systems, $u_c$ is nearly constant at large $\phi$ but
decreases strongly at lower $\phi$ until it vanishes at $\phi_c$.
$\phi_c$ depends on the damping coefficient but approaches recent
estimates of random close-packing \cite{ohern} in the small damping
limit.  In this limit, $u_c$ is determined by the speed of shear waves
in the material, $u_T$.  When $u$ exceeds $u_c=u_T/2$, large shear
strain occurs before shear waves are able to traverse the system. In
the overdamped limit, $u_c$ is nearly independent of $\phi$ over the
studied range and scales linearly with the damping coefficient $b^*$.

%LK question - why does it have to be static friction?  The simulations
%I was doing showed that kinematic friction was sufficient to change
%the profiles (well, just another energy loss mechanism).  Perhaps it
%would be enough to just leave `friction'?  But, if you have some reason
%to emphasize static friction, that's fine with me.

Possible future directions include adding friction to the model
in Eq.~\ref{dissipative}, which would allow us to make more direct
contact with experiments on granular materials \cite{mueth,granular}.
Are nonlinear velocity profiles more likely to occur in systems with 
frictional particles?  How does the propagation of shear waves change
in that case?  We are currently investigating these important
questions.

We thank R. Behringer, A. Liu, and M. Robbins for helpful comments and
one of the referees for suggesting measurements of the shear wave
speed $u_T$.  Financial support from NASA grants NAG3-2377 (NX) and
NNC04GA98G (LK) and Yale University (NX,CSO) is gratefully
acknowledged.


\begin{references}
\bibitem{remark} To simplify the calculations, we measured $u_T(\phi)$
in low-temperature, quiescent reference systems.  This is justified
because the $\phi$ contribution to $u_T$ is much larger than the
fluctuation contribution to $u_T$ for systems with $\phi$ above random close
packing and $u \sim u_c$.
\end{references}
\end{document}